\documentclass[aps,prl,nofootinbib,twocolumn,superscriptaddress]{revtex4}

\usepackage{graphicx}

\newcommand{\beq}{\begin{equation}}
\newcommand{\eeq}{\end{equation}}

\newcommand{\ie}{{\it i.e.}}

\newcommand{\Eq}[1]{Eq.~(\ref{#1})}

\newcommand{\rfn}[1]{(\ref{#1})}
\newcommand{\bea}{\begin{eqnarray}}
\newcommand{\eea}{\end{eqnarray}}

\newcommand{\gsim}{\lower.7ex\hbox{$\;\stackrel{\textstyle>}{\sim}\;$}}
\newcommand{\lsim}{\lower.7ex\hbox{$\;\stackrel{\textstyle<}{\sim}\;$}}

\def\mysection#1{{\bf #1.} }

\begin{document}

\title{Effective axial-vector coupling of gluon as an explanation of the 
top quark asymmetry }

\author{Emidio Gabrielli}
\email[]{emidio.gabrielli@cern.ch}
\author{Martti Raidal}
\email[]{martti.raidal@cern.ch}
\affiliation{NICPB, Ravala 10, 10143 Tallinn, Estonia}

\date{\today}

\begin{abstract}
We explore the possibility that the large  $t\bar t$ forward-backward asymmetry measured by the CDF detector at Tevatron 
could be due to a universal  effective axial-vector coupling of gluon. Using an effective field theory approach we
show model independently how such a log-enhanced coupling occurs at 1-loop level. 
The interference with QCD gluon vector coupling naturally induces
the observed {\it positive} forward-backward $t\bar t$ asymmetry that grows with $t\bar t$ invariant mass and is consistent with the 
cross section measurements. This scenario does not involve new flavor changing couplings nor operators that interfere with QCD,
and, therefore,  is not constrained by the LHC searches for 4-quark contact interactions.
We predict top quark polarization effects that grow with energy and allow to test this scenario at the LHC.
Our proposal offers a viable alternative to new physics scenarios that explain the  $t\bar t$ forward-backward asymmetry anomaly with 
the interference between QCD and tree level new physics amplitudes. 
\end{abstract}


\maketitle

\mysection{Introduction} 
The CDF measurement~\cite{Aaltonen:2011kc} of  large $t\bar t$ forward-backward  (FB) asymmetry $A_{FB}^t=0.445 \pm 0.114$ 
for $t\bar t$ invariant mass $m_{t\bar t} > 450$~GeV came as a surprise. 
It is unexpectedly large compared to the standard model (SM) next-to-leading order prediction 
$A_{FB}^t=0.09 \pm 0.01$~\cite{Bowen:2005ap,Antunano:2007da,Almeida:2008ug}, 
it grows with $t\bar t$ invariant mass since  $A_{FB}^t=-0.116 \pm 0.153$ for  $m_{t\bar t} < 450$~GeV~\cite{Aaltonen:2011kc}, 
and its sign is positive, \ie, opposite to that predicted by the most natural new  physics scenarios with axi-gluons~\cite{axigluon,Ferrario:2008wm,Haisch:2011up}. 
Because of those unusual properties   the numerous specific model dependent~\cite{Rodrigo:2010gm,wang:2011taa}  and model 
independent~\cite{Blum:2011up,Delaunay:2011gv,AguilarSaavedra:2011vw} new physics (NP)
solutions that explain $A_{FB}^t$ with the interference between QCD and tree level NP amplitudes all suffer from similar problems.  
They predict large asymmetry also for $m_{t\bar t}<450$~GeV, large increase of $t\bar t$ cross section at high energies due 
to the QCD interference with NP amplitudes, 
they are strongly constrained~\cite{Bai:2011ed} by the LHC bounds on four-quark contact interactions~\cite{Khachatryan:2011as,ATLAS}, 
and often involve new flavor changing (FC) couplings  to top quark.
Therefore it is not surprising that the latter class of models is already  now strongly disfavored  by 
the LHC~\cite{Collaboration:2011dk}.

In this paper we suggest that, if the $A_{FB}^t$ anomaly with the measured properties persists,  
it could be induced by anomalously large effective axial-vector coupling of the gluon, $g_A,$ that is poorly tested today. 
We show that  NP at the scale $\Lambda\sim1$~TeV can generate large $A_{FB}^t$  with the following properties: 
$(i)$ the sign of  $A_{FB}^t$  is automatically positive; 
$(ii)$ the asymmetry grows with $m_{t\bar t}$ due to the gauge invariance of the axial-vector  form-factor; 
$(iii)$ the asymmetry is enhanced by large logarithm squared;
$(iv)$ no dangerous FC couplings are needed to exist; 
$(v)$ present LHC bounds on four-quark contact interactions do not constrain this scenario; 
$(vi)$ the predicted top quark pair production cross sections at Tevatron and LHC can be consistent with all measurements.
Because  of the latter,
 our most interesting prediction for the LHC experiments
are related to the top quark polarization effects~\cite{Cao:2010nw,Jung:2010yn,Choudhury:2010cd,Krohn:2011tw}
that should be significantly different from the SM predictions. Therefore this scenario can be fully tested by the LHC using those observables.

\mysection{Theoretical framework}
The most general effective Lagrangian for
quark-gluon interactions, compatible with gauge and CP-invariance, 
is 
\bea
{\cal L}&=& -ig_S\Big\{ \bar Q T^a\left[\gamma^\mu\left(1+ g_V(q^2,M)
+ \gamma_5 g_A(q^2,M)\right) G^a_\mu\right. \nonumber \\
&+&
\left.  g_P(q^2,M)q^{\mu}\gamma_5 G^a_\mu+ 
g_M(q^2,M) \sigma^{\mu\nu} G^{a}_{\mu\nu} \right] Q\Big\} ,
\label{vertex}
\eea
where $g_S$ is the strong coupling constant, $G^a_\mu$ is the gluon field, $T^a$ are the color matrices, 
$M$ is the scale of NP, $q^2$ is the invariant momentum-squared carried by the gluon and $Q$ denotes a generic quark field.
At the moment we do not make any assumption on the origin of the form factors
$g_{A,P}(q^2,M)$. In the most general case the form factors $g_{A,P}$ depend also on  quark masses that can be neglected for
$m^2_Q\ll q^2,M^2.$
The last term in Eq.(\ref{vertex}) is the contribution of the chromomagnetic
dipole operator that does not significantly contribute to $A_{FB}^t$~\cite{Blum:2011up}.

Model independently the QCD gauge invariance requires that $2m_Q g_A(q^2,M)=q^2 g_P(q^2,M)$,
thus
\bea
\lim_{q^2\to 0} g_{A,V}(q^2,M)=0\, ,
\label{ginv}
\eea
since no $1/q^2$ singularities are present in $g_P$.
Equation (\ref{ginv}) does not pose any constraint
on the form factors $g_{A}$ and $g_{V}$, which could 
have different magnitudes at arbitrary $q^2$. 
Therefore, gauge invariance does not prevent us to have $g_V\ll g_A$ as long as $q^2\neq 0.$
We stress here once again that the QCD gauge invariance is not broken and gluon remains massless because 
$g_{A}$ and $g_{V}$ are induced via the form factors in \Eq{vertex} 
that are subject to the condition in Eq.(\ref{ginv}).
 Thus the  $g_{V,A}$ exist even in the SM, where they are induced
by electroweak radiative corrections, 
but are numerically 
too small to have significant impact on the observables we consider. 
However, if the origin of large $A_{FB}^t$ is due to NP that has $(V\pm A)$ currents as in the SM, 
large $g_{V}$ and $g_{A}$ can be generated. This is phenomenologically unacceptable because
$g_V$ is strongly constrained by the total $q\bar q\to t\bar t$ cross section
that depends quadratically on $g_A$ but only linearly on $g_V$.
Therefore, from now on, we will neglect the contribution of the vectorial 
form factor $g_V(q^2,M)$ in Eq.(\ref{vertex}), and consider only NP scenarios that
generate $g_A$ with  the hierarchy $g_V \ll g_A$.

In the limit of $q^2\ll M^2$, it is useful to parametrize the 
axial-vector form factor
as
\bea
g_A(q^2,M)=\frac{q^2}{\Lambda^2} F(q^2,\Lambda)\, ,
\label{gA}
\eea
where  we absorb the NP coupling $\alpha_{NP}$ and loop factor into the NP scale, $\Lambda^2=M^2/(4\pi \alpha_{NP}).$
Because of the breaking of conformal invariance, induced by renormalization, 
we expect \cite{Raidal:1997hq} $F(q^2,\Lambda)$ to contain also logarithm terms  $\log(q^2/\Lambda^2).$
This could give a large log enhancement in the case of $|q^2|\ll \Lambda^2$.
In general, the form factor $F(q^2,\Lambda)$ could also 
develop an imaginary part for $q^2>0$.
In perturbation theory, this 
is related to the absorptive part of the loop diagram generating $g_A$,
when $|q^2|$ is above the threshold of some specific particles pair production.

\mysection{$A_{FB}^t$ and $t\bar t$ cross section}
We first show that the large $A_{FB}^t$ can be generated consistently with the $t\bar t$ cross section constraints
via the operator
 \bea
g_S \frac{q^2}{\Lambda^2} F(\frac{q^2}{\Lambda^2}) 
[\bar Q \gamma^\mu\gamma_5  T^a Q] G^a_{\mu}.
 \label{op}
 \eea
The $t\bar t$ FB asymmetry occurs  due to the interference between the gluon mediated s-channel SM amplitude for
$q\bar q\to g \to t\bar t$
 and the analogous s-channel amplitude induced by two vertices of \Eq{op}. First, the induced asymmetry
 grows with the invariant mass of the $t \bar t$ system $q^2=s=(p_t+p_{\bar t})^2$ exactly as observed.  Second, the sign for the
 asymmetry comes out to be the right one due to the massless gluon, provided 
the sign of the form factor $F(q^2/\Lambda^2)$ is universal. 
Third,  it is expected to give only a subdominant
  contribution to the $t\bar t$ production cross section that agrees very well with the SM predictions.
 In fact, it is predicted~\cite{Stephenson} 
that the only observable where the operator in 
\Eq{op} could play a dominant role
 is the $\bar QQ$ asymmetry at very high energies, and this is exactly 
what is happening.
The question to address now is how large form factors $F$ are needed to explain the observed $t\bar t$ asymmetry and whether
there exist new physics models that can induce it naturally.

We have evaluated the FB asymmetry and total cross section of $t\bar t $
production at  the leading order (LO) in QCD by including the contribution
of the axial-vector coupling in Eq.(\ref{vertex}).
In the total cross section, we have also included 
the contribution due to the gluon-gluon 
partonic process  $gg\to t\bar t$. However, since this last process 
is subleading at Tevatron,
giving only a 10\% effect of the total cross section, we have retained
in the $gg\to t\bar t$ amplitude only the SM contribution.

By using the notation of Eq.(\ref{vertex}), 
the partonic total cross section
for $q\bar{q}\to t \bar{t}$, in the limit of $m_q\to 0$, is given by
\bea
\sigma_{q\bar q}^+(\hat s) - \sigma_{q\bar q}^-(\hat s)&=&
\frac{8\pi\alpha^2_S \beta^2_t}{9 \hat{s}} {\rm Re}[g_A^t]{\rm Re}[g_A^q],
\label{afb}
\\
\sigma_{q\bar q}^+(\hat s) + \sigma_{q\bar q}^-(\hat s)&=&
\frac{8\pi\alpha^2_S \beta_t}{27 \hat{s}}\Big\{(1+2\frac{m_t^2}{\hat s})
\left(1+|g_A^q|^2\right)+
\nonumber \\
&&
\beta_t^2 |g_A^t|^2\left(1+|g_A^q|^2\right)\Big\} ,
\label{xsections}
\eea
where $g_A^t$ and $g_A^u$ are the effective
axial-vector gluon couplings of the top- and $q$-quarks respectively, 
$\beta_t=\sqrt{1-4m_t^2/\hat{s}}$ is the top-quark velocity in the 
$t\bar t$ rest frame, $\hat{s}=x_1x_2 S$ with $\sqrt{S}$ the $p \bar{p}$
center of mass energy, and $x_{1,2}$
the corresponding fractions of partons momenta.
Here $\sigma^{\pm}$ indicate the inclusive cross section integrated 
in the positive/negative range of $\cos{\theta}$ respectively, 
where $\theta$ is the angle between the direction of the outgoing top-quark
and the initial quark momentum. 

Then, after the convolution of parton cross
sections in Eq.(\ref{xsections}) with parton distribution functions,
the $t\bar t$ FB asymmetry at the LO is given by
\bea
A_{FB}=\frac{\sum_q\int d \mu_q
\left(\sigma^+_{q\bar q}(\hat s)-\sigma^-_{q\hat q}(\hat s)\right)}
{\int \left( \sum_qd \mu_q \sigma_{qq}(\hat s)+
d\mu_{g}\sigma_{gg}(\hat s)\right)},
\label{AFB}
\eea
where $\sigma_{qq}=\sigma_{q\bar q}^+ + \sigma_{q\bar q}^-$, 
$\sigma_{gg}$ the $gg \to q\bar q$ cross section, 
while 
$d \mu_{q}$ and $d \mu_{g}$ indicate the differential integrations in 
$dx_1 dx_2$ convoluted with the 
quarks and gluon parton distribution functions respectively.
The $\sigma_{gg}(\hat s)$ stands for the 
total cross section of $gg\to t\bar{t}$ at the LO, whose
SM analytical expression can be found in \cite{Beenakker}.

We estimate now how large the form factors we need to explain the large FB
asymmetry observed.
We consider the most simple case assuming $g_A^t=g_A^u$ and
$F(q^2/\Lambda^2)=1,$   
that is neglecting any possible enhancement from large $\log(q^2/\Lambda^2)$ terms.
This is a good approximation in case one is interested in
 estimating the scale $\Lambda$
that could provide the $A^t_{FB}$ asymmetry.

\begin{figure}[t]
\begin{center}
\includegraphics[width=0.33\textwidth, angle=-90]{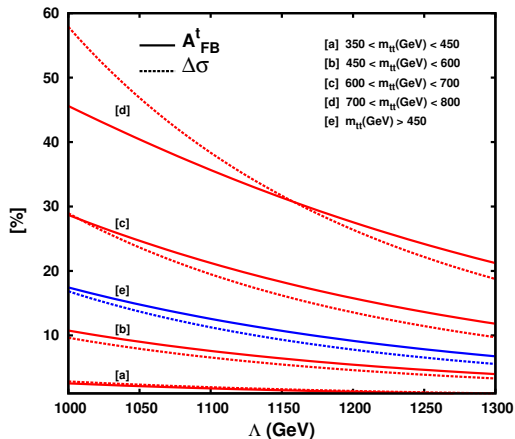}
\vspace{-0.3cm}
\caption{The $t\bar t$ FB asymmetry $A_{FB}^t$ (continuous lines) and 
the cross section variation $\Delta \sigma\equiv \sigma/\sigma_{SM}-1$ 
(dashed lines),  as a function of the scale $\Lambda$, for 
several ranges of integrations in the $m_{tt}$ invariant mass.
} 
\label{fig2}
\end{center}
\end{figure}

In the numerical integration of Eq.(\ref{AFB}) we have used the 
CTEQ6L1 parton distribution functions (PDF) \cite{PDF} and the total cross
sections evaluated at the LO in QCD. We set the PDF scale $\mu=m_t$ with 
top-quark mass $m_t=172$ GeV.
We present our results in Fig.\ref{fig2}, where we have
plotted the new physics contribution to FB asymmetry (continuous lines) 
and the cross section variation defined as 
$\Delta \sigma\equiv \sigma/\sigma_{SM}-1$ (dashed lines), 
versus the scale $\Lambda$, 
for several regions of integrations in the $m_{tt}$ invariant mass.
The results for the FB asymmetry $A^t_{FB}$ in Fig.\ref{fig2} are 
not very sensitive to the choice of the PDF scale. Moreover, we 
also expect that the inclusion of the next-to-leading order QCD corrections 
will not change dramatically these predictions, due to the 
expected factorization property of QCD corrections
to cross sections $\sigma(\pm)$ in Eqs.(\ref{afb})-(\ref{xsections}).

We plot the result for the range 
$1 {\rm TeV} < \Lambda < 1.3 {\rm TeV}$, 
where the asymmetry $A^t_{FB}$ for
$m_{t\bar t}>450$ GeV is larger than SM value and the maximum variation
of total cross section is below 20\%.
In Fig.\ref{fig2}, the red curves indicated with [a],[b],[c],[d],
correspond to the bins of $m_{t\bar t}$ integrated in the ranges
$(350-450)$~GeV, $(450-600)$~GeV, $(600-700)$~GeV,
$(700-800)$~GeV, respectively, while the curves [e] stand for 
$m_{t \bar t}> 450$~GeV.
Notice that results in Fig.\ref{fig2} do not include the SM contribution to $A^t_{FB}$ that further increases the signal.

The main trend of this scenario  is characterized by a 
FB asymmetry that grows  with $m_{t\bar t}$ invariant mass.
This is expected from the fact that the $g_A$ form factor is proportional to 
$q^2/\Lambda^2$. We stress that our scenario is in perfect agreement with data for low invariant masses, bin [a], 
while for the last bin [d] we have $A^t_{FB}\sim 45\%$ at $\Lambda=1$ TeV.
However, a large FB asymmetry also implies a large contribution to the total
LO cross section, due to the constructive interference 
between the SM amplitude and the one with axial-vector gluon couplings
in the s-gluon channel.
Still, for the bins [a,b,c] that dominate statistically, the variation of the total 
cross-section $\Delta \sigma$ remains small, below the $30\%$ level. 
The same is true for the observables for $m_{t\bar t}> 450$ GeV, shown in the region 
[e]. Those results are consistent with the total inclusive asymmetry~\cite{Jung:2009pi,Degrande:2010kt}
and with cross section measurements since the characteristic uncertainty
associated to cross section in the lowest bins can be still of the order 
of $30-40 \%$. In the last bin of integration [d],
$\Delta \sigma$ tends to be larger than $30\%$ for $\Lambda < 1.2$ TeV.
However, we should take into account that the experimental uncertainty 
of the cross section for the bin [d] is larger than the corresponding ones at 
lower values of $m_{t\bar t}$, due to the lack of statistics at high 
$m_{t\bar t}$.

The fact that we need a low-energy scale $\Lambda\sim 1-1.3$~TeV to generate
a large contribution to the FB asymmetry of order $A^t_{FB}\sim 10-20$\% for 
$m_{t\bar t} > 450$~GeV suggests that in the most general case the two
scales $\Lambda_t$ and $\Lambda_q$, related to the top-quark and light-quark 
vertex respectively, should be comparable.
Indeed, in this case the magnitude of the asymmetry is controlled by 
the geometric average of the two scales, namely 
$\Lambda \sim 
\sqrt{\Lambda_t\Lambda_q}$. If we need $\Lambda$ to be of order of 
1 TeV, we cannot push $\Lambda_q$ too high, since $\Lambda_t$
would be close to the EW scale and the contribution to 
total cross section would explode. Therefore, results in Fig.\ref{fig2}
suggest that the two scales should be comparable, supporting the idea of 
a universal $g_A$ coupling.

\begin{figure}[t]
\begin{center}
\includegraphics[width=0.22\textwidth]{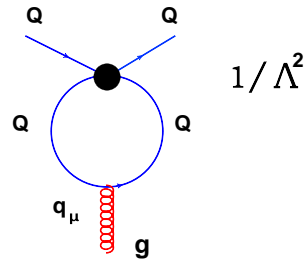}
\vspace{-0.3cm}
\caption{Feynman diagram in the effective low-energy theory
that generates the effective axial-vector coupling of gluon.} 
\label{fig1}
\end{center}
\end{figure}

\mysection{The origin and constraints}
Assuming that NP is perturbative, model independently the effective operators~\cite{Delaunay:2011gv}
\bea
O^{1,8}_{AV} &=&\frac{1}{{\Lambda}^2}
[\bar Q T_{1,8}\gamma^\mu\gamma_5  Q] [\bar Q T_{1,8} \gamma^\mu Q]\, , 
\label{OAV} \\
O^{1,8}_{PS}&=&\frac{1}{{\Lambda}^2}
[\bar Q T_{1,8}\gamma_5  Q] [\bar Q T_{1,8} Q] ,
\label{OPS}
\eea
 generate $g_A$ via 1-loop diagrams depicted in Fig.~\ref{fig1}.
Here $T_1=1$ and $ T_8=T^a,$ thus both isoscalar and octet operators contribute. Notice that:
 $(i)$ no $g_V$ is induced due to QCD parity conservation;
   $(ii)$ the 1-loop induced $g_A$ is enhanced by $\log(q^2/\Lambda^2)$;
 $(iii)$ the operators $O^{1,8}_{AV}$, $O^{1,8}_{PS}$ do not induce FC processes; however, there could be different quark flavors in the loop
 in Fig.~\ref{fig1}, extending the operator basis to $Q\to Q',$ $V\leftrightarrow A,$ $P\leftrightarrow S$ is straightforward; 
 $(iv)$  the operators $O^{1,8}_{AV}$, $O^{1,8}_{PS}$ do not  interfere with the corresponding QCD induced 4-quark processes.
 The latter point has very important implications for our scenario -- the stringent LHC constraints~\cite{Khachatryan:2011as,ATLAS} on
 4-quark contact interactions do not apply at all. Indeed, those constraints come from the interference between QCD and NP diagrams, and
 constrain the  models that  explain $A^t_{FB}$ with the similar interference very stringently. We stress that our scenario is free from those constraints
 and NP at 1-2~TeV can induce large $g_A$ as explained above.

Alternatively, large $g_A$ might be generated by new strongly-coupled
parity-violating dynamics related to electro-weak symmetry breaking 
(EWSB) at 1-2 TeV scale.
Because this NP is entirely nonperturbative,  generating $g_A$ is possible~\cite{Lane:1996gr}
but we are not able to compute it. We are only able to estimate 
the validity range of the effective coupling parametrization in Eq.(\ref{gA}) that is controlled by
$\hat{s}/\Lambda^2_{eff}$, where $\Lambda^2_{eff}$  is 
expected to be related to $\Lambda$ as 
$\Lambda_{eff} \sim \Lambda/\sqrt{\alpha_S}$. 
For $\Lambda \sim 1 (1.3)$ TeV, as required by the 
$A^t_{FB}$ anomaly, the related scale is 3.5 (4.6) TeV. 
At this scale a plethora of new resonances should occur at the LHC allowing to test this scenario. 
Notice that, in the region of large invariant masses 
$\hat{s}\gg \Lambda_{eff}^2$, the low-energy ansatz
$g_A\sim q^2/\Lambda^2$ is not valid anymore and the $q^2$
dependence of $g_A$ should be determined by fitting the data.

\mysection{Observables at LHC and other experiments}
First, the natural question to be asked is  whether our solution to the Tevatron $t\bar t$ asymmetry is related to the $b\bar b$ $A_{FB}$ anomaly observed at
$Z$ pole at LEP. Obviously the answer is no because the initial state at LEP is $e^+e^-$ and gluon does not couple to leptons. 
In addition, the induced form factors do not affect precision data involving QCD at observable level. Firstly the form factors vanish at low energies according to 
\Eq{gA}. Second, at high energies the loop induced $g_A$ is still much smaller than unity and, taking into account relatively large experimental
 errors in the QCD measurements, does not affect QCD observables. 

As we stressed before, the observables that are sensitive to $g_A$
 are asymmetries.
Because the anomalous axial-vector coupling of gluon grows with $q^2$, see \Eq{op}, the most natural test of our scenario 
is measuring $t\bar t$ cross section dependence on $m_{t\bar t}.$ However, at the LHC the dominant production process 
$gg\to t \bar t$ is induced by the s-channel diagram with 3-gluon vertex in addition to the t,u-channel diagrams . The $g_A$ coupling is expected to affect the s-channel diagram since there the gluon coupled to fermions is off-shell
\cite{us}. This is a peculiar prediction of our scenario which differs from most popular axigluon models where the  $gg\to t \bar t$ production mechanism is not affected by the new physics. Moreover, our scenario is testable at the LHC experiments because $g_A$ coupling should induce observable polarization effects of top quarks that also grow  with $m_{t\bar t}.$
Therefore, studies of top quark polarization at the LHC
are sensitive tests of our scenario as well as other models of physics beyond the SM~\cite{Cao:2010nw,Jung:2010yn,Choudhury:2010cd,Krohn:2011tw}.
A dedicated LHC study is needed to discriminate between different sources of asymmetries and polarizations.

\mysection{Conclusions}
Among many model dependent and model independent solutions proposed to explain the measured top quark FB asymmetry, 
our proposal is the only one that  does not involve interference between the QCD and tree level NP contributions mediated by heavy resonances. 
Instead, we argue that the large 
$A_{FB}^t$ is induced by an anomalously large effective axial-vector coupling of the gluon, $g_A$, described at low energies by the operator in \Eq{op}.
We have shown that $g_A$ can explain the sign, the magnitude and the behavior of  $A_{FB}^t$ consistently with the $t\bar t$ cross section measurements.
We have shown model independently that logarithmically enhanced  $g_A$ can be induced by NP effective operators \rfn{OAV}, \rfn{OPS} that 
do not suffer from flavor constraints and from the LHC constraints on 4-quark contact interactions.
While our results are presented in the context of top quark FB asymmetry,
our proposal to study anomalous axial-vector coupling of gluon has physics implications beyond that observable.
Studying the induced top quark cross sections, distributions and polarization effects at the LHC 
allows one to test different classes of models beyond the SM.

\vskip 0.4cm

\paragraph{Acknowledgement.}
We thank the Les Houches 2011 BSM working group members who helped to formulate this study, 
in particular C. Delaunay and R. Godoble for discussions on  top quark asymmetry and polarization, and B. Mele and G. Rodrigo for several discussions.
This work was supported by the Estonian Science Foundation under Grants Nos.  8090 and MTT60, and by the Estonian Ministry of  Education under the SF0690030s09 project.

\end{document}